# GRRIS: A Real-time Intrasite Observation Scheduling Scheme for Distributed Survey Telescope Arrays

Yajie Zhang[1,2], Ce Yu[1,2], Chao Sun[1,2], Yi Hu[3], Zhaohui Shang[3], Jizeng Wei[1,2], and Xu Yang[3]
[1] College of Intelligence and Computing, Tianjin University, No. 135 Yaguan Road, Haihe Education Park, Tianjin 300350, People's Republic of China
[2] Technical R&D Innovation Center, National Astronomical Data Center, No. 135 Yaguan Road, Haihe Education Park, Tianjin 300350, People's Republic of China
[3] National Astronomical Observatories, Chinese Academy of Sciences, No. 20 Datun Road, Chaoyang District, Beijing 100012, People's Republic of China

## Abstract

The distributed telescope array offers promise for conducting large-sky-area, high-frequency time-domain surveys. Multiple telescopes can be deployed at each observation site, so intrasite observation task scheduling is crucial for enhancing observation efficiency and quality. Efficient use of observable time and rapid response to special situations are critical to maximize scientific discovery in time-domain surveys. Besides, the competing scientific priorities, time-varying observation conditions, and capabilities of observation equipment, lead to a vast search space of the scheduling. So with the increasing number of telescopes and observation fields, balancing computational time with solution quality in observation scheduling poses a significant challenge. Informed by the seminal contributions of earlier studies on a multilevel scheduling model and global scheduler for a time-domain telescope array, this study is devoted to further exploring the site scheduler. Formulating the observation scheduling of multiple telescopes at the site as a cooperative decision-making problem, this paper proposes GRRIS, a real-time intrasite observation scheduling scheme for the telescope array using graph and reinforcement learning (RL). It employs a graph neural network to learn node features that can embed the spatial structure of the observation scheduling. An algorithm based on multi-agent RL is designed to efficiently learn the optimum allocation policy of telescope agents to field nodes. Through numerical simulations with real-world scenarios, GRRIS can achieve up to a 22% solution improvement over the most competitive scheme. It offers better scalability and subsecond decision speed, meeting the needs of observation scheduling control for future distributed telescope arrays.

*Unified Astronomy Thesaurus concepts:* Surveys (1671); Astronomical models (86); Astronomical methods (1043); Observational astronomy (1145)

## 1. Introduction

As the primary form of exploration of the Universe by astronomers, astronomical observation has produced a large number of scientific discoveries (e.g., T. Santana-Ros et al. 2022; T. Saifollahi et al. 2023). In recent decades, the development of astronomy has entered a new phase of comprehensive exploration of the dynamic Universe, as the rapid changes of celestial objects have attracted widespread attention. Wide-field, multiband, and high-cadence time-domain sky surveys are the inevitable trend (Y. Zhang et al. 2023). This is beyond the capability of a single telescope design, so the efficient combination of multiple telescopes in a telescope array has the potential to maintain both the observation coverage area and observation frequency (J. Liu et al. 2021). In practice, efficient scheduling of observation time is essential to maximize scientific output, due to the time-varying visibility of celestial objects and the expensive and limited lifetime of astronomical observation facilities.

Various telescope observation control and scheduling schemes have been proposed to improve observation efficiency and scientific output. However, studies of the scheduling methods for Vera C. Rubin Observatory[4] (E. Naghib et al. 2019) and Zwicky Transient Facility (ZTF; E. C. Bellm et al. 2019), for example, are based on the allocation of observation time to the single telescope and lack the consideration of the interactions between different telescopes in a telescope array. Moreover, the Las Cumbres Observatory Global Telescope (LCOGT), a robotic telescope network, implements similar integer linear programming (ILP) techniques to that of ZTF to assign requested observations to telescopes. The scheduling model of LCOGT is a proposal-based scheduling, the scientific value of which is assessed by the Time Allocation Committee. This is different from the scheduling scenario of multiple geographically distributed telescope arrays to complete time-domain surveys, which is studied in this paper. In addition, heuristic methods are widely used for solving astronomical observation scheduling problems, e.g., adjusting exposure time dynamically according to airmass (J. Rana et al. 2019), schedule by the minimum telescope slew angle (J. Rana et al. 2017), etc., for finding optical counterparts of the transients. However, for practical large-scale survey observations, determining when and where to apply these heuristics, and prioritizing them, can be inflexible and challenging due to the complexity of scheduling, the variability of observational conditions, and the need for timely decision-making.

For the telescope array, in addition to considering the priority constraint, astronomical observation conditions, telescope performance, quality feedback, etc., it also needs to consider the potential observation time and resource competition between telescopes. Many of these constraints are subject to change due to special conditions, requiring telescopes to collaborate, which can pose challenges in implementing observation plans. Currently available methods for scheduling telescope arrays are based on small-scale surveys and still require manual intervention (F. G. Saturni et al. 2022). Most existing telescope array scheduling methods are designed based on problem-specific operational research methods and heuristics (Q. Liu et al. 2018; B. Parazin et al. 2022; J. Sun et al. 2022), just to name a few). They often require significant expertize, and as the number of telescopes and observation fields increases, the computational complexity of the scheduling will increase exponentially. So the high computation complexity prevents these methods from being widely applied to real-world distributed large-scale problems. Besides, solving constrained optimization problems is usually time-consuming, posing challenges in responding to rapidly changing constraints and emerging transient targets.

---

[4] See https://project.lsst.org/.

In recent years, encouraged by the development of reinforcement learning (RL), neural-based methods have been developed to solve various traditional combinatorial optimization (e.g., L. L. Tong et al. 2022; H. Gao et al. 2023) and practical scheduling problems, such as electric vehicle charging (S. Wang et al. 2019), resource scheduling in smart factories (e.g., L. Chen et al. 2022; Z. Zhang et al. 2023), multi-unmanned aerial vehicle networks (S. Khairy et al. 2020), and Earth observation satellite scheduling (X. Wang et al. 2020), which have shown promising advantages over heuristic algorithms. And yet, such approaches are conducted rarely in the astronomical observation domain, possibly because there are many difficulties involved in designing an efficient optimization algorithm: (1) compared with general scheduling problems, astronomical observation scheduling is a continuous long-term process, the constraints and optimization objectives need to consider not only the spatial characteristics of the observation targets, but also more prominent temporal computational characteristics; (2) efficient architectures are needed to capture the implicit influence between telescopes that construct their own observation task queues for common goals; (3) as anondeterministic polynomial-time hard problem (R. Bonvallet et al. 2010), it is expensive to solve the telescope observation task queue, so there is a lack of data for the optimal solution of the ground truth.

F. Terranova et al. (2023) implement a scheduling framework for self-driving telescopes using offline RL. Results suggest that RL algorithms can be used to optimize the sequential schedule of a telescope survey. Moreover, P. Jia et al. (2023) develops a telescope array observation simulator and applies deep reinforcement learning (DRL) to a space debris observation scenario. Therefore, extensive research on RL as an efficient training autonomous system provides the basis for optimizing the observation scheduling scheme of the time-domain survey telescope array.

We propose a two-step approach, *global scheduler* and *site scheduler*, based on the multilevel scheduling model (Y. Zhang et al. 2023). The global scheduler is designed as a long-term scheduler that acts as a central node to coordinate the telescopes at all observation sites and globally controls the progress of the survey and the utilization of resources. As shown in Figure 1, the global scheduler calculates the available long-term scheduling blocks for each site during a long time coverage, and then hands them to the site schedulers to assign specific observation tasks to the telescopes. The time slots of the scheduling blocks for different sites are different due to the different site locations and observation conditions.

This paper is about the site scheduler, which aims to further study the optimization of multi-telescope observation in a single site. Notably, due to Earth's rotation, the observation quality of a task at a fixed site varies with time, particularly concerning the altitude angle and airmass. We introduce the graph neural network (GNN) to embed the working graph of cooperative optimization problems into latent spaces, incorporating actual telescope locations, celestial object distributions, and dynamic astronomical observation conditions. It can offer advantages in terms of spatial structure representation and scalability for larger-scale problems. We also develop a multi-agent reinforcement learning (MARL) algorithm with an attention mechanism that is able to analyze the embedded graph, and make decisions assigning agents to the different vertices. The contributions of this paper are threefold.

(1) An efficient intrasite observation scheduling strategy is proposed for a distributed multi-telescope-enabled observation system, where the telescopes in the observation site complete the observation tasks in the scheduling block collaboratively. The scheduling optimization problem is formulated to obtain the maximum total observation quality and minimum computational time considering time-varying observation conditions and the capability of the telescopes.
(2) By formulating the highly complex nonconvex optimization problem as a sequential decision-making problem, we propose GRRIS, a graph-based RL scheme to learn an intrasite observation scheduling policy representation. The attention mechanism is used to allow telescope agents to make individual decisions. An observation quality reward is designed to reflect the immediate payoff of any decision in all decision steps.
(3) By simulations performed on real-world data, GRRIS is shown to outperform five representative scheduling optimization methods, especially in terms of the decision speed, robustness to the change of observation time and geographical distributions, and scalability for increasing observation scale. It can also learn a transferable scheduling policy that can be employed for various numbers of observation tasks without further training.

The remainder of this paper is organized as follows. Section 2 presents the real-life astronomical observation background and problem formulation. Section 3 casts the observation scheduling problem into the MARL framework and describes the proposed GNN and distributed policy network solutions. Section 4 presents the simulation results, and Section 5 discusses future challenges and draws conclusions.

## 2. Telescope Array Observation Scheduling Preliminary and Problem Formulation

### 2.1. Preliminary Concepts

A large-scale telescope array is distributed across multiple observation sites, each hosting multiple telescopes managed by site owners for operational control. Based on the multilevel scheduling model of the telescope array proposed in Y. Zhang et al. (2023), we assume long-term scheduling results for each site are available, requiring fine-tuning of individual telescope scheduling based on facility status and time-varying observation quality factors. Each telescope within the site is capable of being controlled in a distributed manner so that it is natural to be considered an agent.

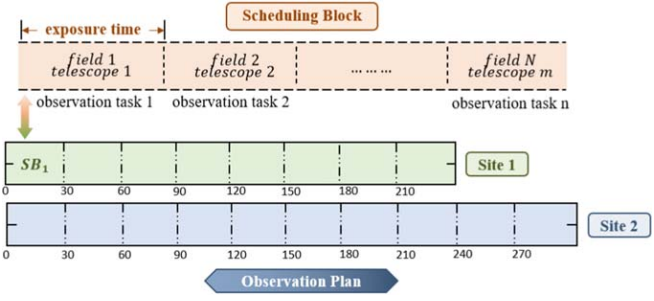

**Figure 1.** Schematic diagram of the scheduling block. The scheduling block refers to the long-term global scheduler output, including a series of fields that each site needs to observe in a long period and their observation time, which are constrained by observation conditions and survey feedback. The scheduling blocks also serve as the input of the site scheduler in the multilevel scheduling model for the telescope array survey scheduling model (Y. Zhang et al. 2023).

As illustrated in Figure 1, the observation plan generated by the global scheduler consists of multiple scheduling blocks, which are computed based on astronomical observation conditions (e.g., moon phase, Sun altitude) at different sites for the observation field and the observation site's capacity. Each observation field is identified by celestial coordinates (R. A., decl.). Scheduling blocks are then assigned to the site scheduler for more refined task allocation. Moreover, the observation time within each scheduling block is partitioned into smaller time slots, aligning with the telescope's exposure time. Here, we assume consistent exposure times for all telescopes at a given site. Each exposure of a specific sky region constitutes an observation task, and our emphasis lies in efficiently assigning these tasks to the telescopes for optimal productivity.

We use airmass as an example to quantify the completion quality of scheduled observation tasks. Airmass, which depends on the zenith angle, reflects the amount of air that starlight must travel through before reaching the telescope. Higher airmass values correspond to more atmospheric interference, resulting in larger extinction in the observed starlight and reducing the quality of observations. Airmass is a straightforward metric that is useful for developing and evaluating scheduling algorithms for demonstration. In reality, we will use a more complicated parameter, "nominal extinction" (Y. Hu et al. 2024, in preparation), to quantify the completion of the observation quality. This parameter can take into account the effects of the moon phase, angular distance of the moon, atmospheric quality, sky brightness, cloud cover, etc., and combine their impacts into a single extinction value in magnitude. Figure 2 demonstrates how airmass changes with time at the same site, considering observation altitude angle and airmass. It can be seen that the observation quality of observation tasks varies with time. Lower-airmass observations are preferred in astronomy for better image quality and less atmospheric distortion (J. Rana et al. 2019). Hence, an efficient intrasite scheduling scheme is required to allocate site-level observation tasks to specific telescopes at optimal observation times, maximizing overall observation quality.

### 2.2. Problem Formulation

In this paper, we model the problem on a directed graph $G = (V, E)$, where $V = \{v_1,...,v_n\}$ is a set of $n$ nodes (fields to be observed in the scheduling block), and $E$ is a set of edges, as shown in Figure 3. The time of the scheduling block will be divided into $N$ exposures according to telescopes' observation capability. So, we denote a set of exposures by $T = \{t_1,...,t_N\}$ for each telescope. All $m$ agents (telescopes) in the site cooperate to complete the observation tasks of $n$ fields in a certain period of time, during which the fields can be observed multiple times. Each observation task must be executed once by any agent. So, the solution to the telescope array scheduling problem is defined as $\pi = \{\pi^1,..., \pi^m\}$, which consists of a set of agent observation tasks. An ordered set of the fields observed by each agent is denoted as $\pi^i = (\pi^i_1, \pi^i_2,..., \pi^i_{n_i})$.

In the scheduling model, the nominal extinction of an agent's observation task is defined as the *observation cost*, computed in this paper based on the airmass of each astronomical observation task (as described in Section 2.1). It is worth noting that the evaluation of the observation cost can be tailored to various astronomical observation scenarios. Between exposures, the telescope transitions to the next sky area. Earth's rotation during this transition can impact the quality of subsequent tasks. So the cost of the observation task of field $v$ by telescope $m$ at time $t$ can be denoted by $c(t, p_a, p_v)$. Here, $p_a$ represents the position of the telescope (agent) $a$, and $p_v$ represents the position of the field $v$. Therefore, the observation cost of agent $i$ for a total of $T_i$ exposures of field $v$ can be described as

$$L(\pi^i) = \sum_{j=1}^{T_i} c_{\pi^i_j}(t_j, p_i, p_v). \quad (1)$$

Here, we consider MinMax (minimizing the max cost among telescope agents) as the objective (i.e., makespan) of our site scheduling model:

$$\text{minimize} \max_{i \in \{1,..., m\}} (L(\pi^i)). \quad (2)$$

In practical astronomical observation applications, the choice of this objective can encourage agents to achieve a globally optimal observation effect collaboratively. In addition, the MinMax method contributes to load balancing among agents since agents with a large number of observation tasks can generate greater airmass, which will be further optimized. The problem necessitates optimizing task allocation, along with the creation of individual task queues. This results in a high-dimensional search space due to numerous time-varying constraints, such as telescope and field locations, exposure times, and more.

## 3. Real-time Intrasite Observation Scheduling Scheme for a Telescope Array

### 3.1. Telescope Array Observation as an RL Problem

By formulating the telescope observation scheduling problem into a MARL framework, we can consider it as a sequential decision-making problem. Each agent performs a sequence of decisions, selecting the next task and interacting with the environment. Agents make asynchronous decisions based on their own observations as they progress to the next field in the observation plan. This collaborative approach can construct a global solution. Each decision can be regarded as an RL action. To account for the time needed for the agent to slew to the next field, we introduce a decision time gap between decision-making steps. When an agent picks a field at a particular time step, the decision time gap is determined by averaging the observation cost between the current and the next field to be observed. The decision time gap decreases by one unit per time step until zero, signifying the agent's arrival at the next observation field and readiness for the next decision. This mechanism encourages asynchronous distributed decision-making among agents, mitigating potential conflicts in field selection.

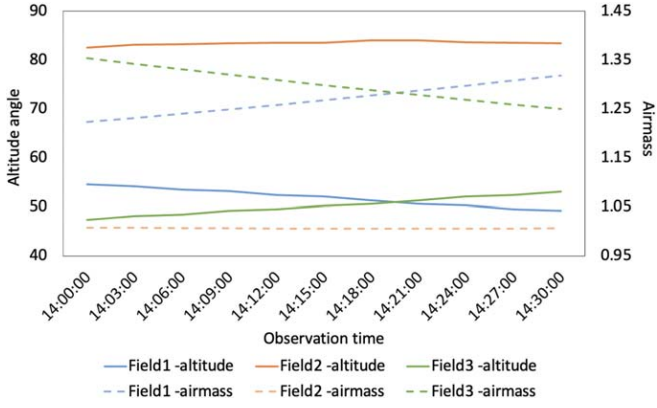

**Figure 2.** Examples of time-varying observation altitude and quality for observations of Field 1 (R.A. 23°, decl. 36°), Field 2 (R.A. 73°, decl. 36°), and Field 3 (R.A. 123°, decl. 36°) starting at 2:00 PM on 2021 January 1, at a site with latitude 42°, longitude 117°, and an elevation of 950 m.

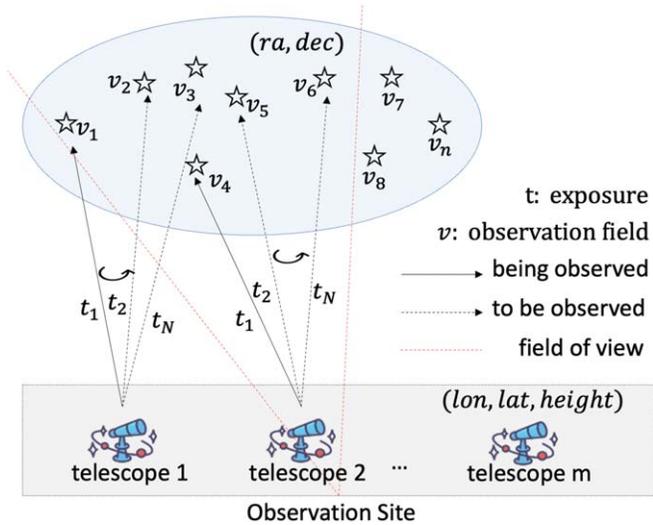

**Figure 3.** Illustration on the telescope observation task execution in an observation site. All $m$ agents (telescopes) in the site cooperate to complete the observation tasks of $n$ fields in the scheduling block. A set of exposures $T = \{t_1,...,t_N\}$ represents the exposures of each telescope.

By framing the site observation scheduling as a sequential decision-making problem, we build the overall model as depicted in Figure 4. In general, the policy starts by summarizing the current state of the graph $G$, then it proceeds to assign nodes to each agent for astronomical observation. Each agent uses its local network to select the next field based on its own observation, decomposing the problem into smaller scheduling sub-problems. To train the policy, we utilize GNN to embed the graph $G$, and create a set of distributed policy networks for node allocation to agents. In the decoder, we use updated field features and the agent's current state to assign attention weights to each field, which serve as its policy.

### 3.2. Graph Embedding

Inspired by the compositional message-passing neural network (CMPNN) proposed in Z. Zhang & W. S. Lee (2019), the node embedding can be updated according to the graph structure. So, the GNN will calculate a $p$-dimensional feature embedding $f_v$ for each node $v \in V$. At each iteration, the CMPNN can be formulated as

$$f_u^{t+1} = \Phi_{v \in \mathcal{N}_u} h_e(e_{uv}) k f_v^t. \quad (3)$$

We use $\mathcal{N}_u$ to represent all neighbors of node $u$, while $\Phi$ for the activation function. A neural network $h_e$ is leveraged to generate an attention vector that predicts the edge types. Then, $k \in \mathbb{R}^{t_e \times d_{in} \times d_{out}}$ can be seen as the global invariant kernel. Through several rounds of message passing, information is propagated to more distant nodes in the graph. If the GNN terminates after $T$ iterations, the final node embedding $f_u^T$ will incorporate information from its $T$-hop neighborhoods.

To parameterize the GNN using the CMPNN framework, the update process of node embedding is designed as

$$f_u^{t+1} = \text{relu}\left\{\max_{v \in \mathcal{N}_u}[\theta_e e_{uv}(\theta_1 f_u^t + \theta_2 f_v^t - \theta_2 f_u^t)]\right\}, \quad (4)$$

where $\theta_1$, $\theta_2$, and $\theta_e$ are the model parameters. $\theta_1$ and $\theta_2$ are shared parameters by all nodes, while $\theta_e$ is shared by all edges. Here, we aggregate the information from neighboring nodes by taking the maximum over their embeddings. After computing the node embeddings for $T$ iterations, we can leverage the embedded node features and graph features to formulate the distributed policy networks.

### 3.3. Agent Embedding

Based on the global information and node embeddings in the graph, each agent independently generates its own agent embedding. Different agents adopt the same process but utilize different parameters. In the computation process of the embedding of an agent, first a set of node features from the graph embedding that are formalized as $f = \{f_1, f_2,...,f_n\}$, and a graph embedding $g_f$ are input. The $g_f$ is computed by max pooling from the set of node features ($f_i$, $g_f \in \mathbb{R}^p$, and $p$ for the dimension of the node embedding). It is defined as $g_f = \max\{f_1, f_2,...,f_n\}$, where the max is applied element-wise to its input. Here, the agent embedding leverages the attention mechanism to output attention coefficients, which indicates the importance of node $i$'s feature for constructing its embedding.

After that, using its own embedding and all agent embeddings, each node independently assigns a telescope agent to itself. In our design, besides the process of agent embedding, we also use the attention mechanism for agent assignment by the nodes. To be more precise, the attention mechanism is adopted for each agent to compute the importance of the node to the agent. As shown in Equation (5), the keys and values are from the set of node features, while the query for the agent $i$ is computed from the global embedding, which is common for all agents.

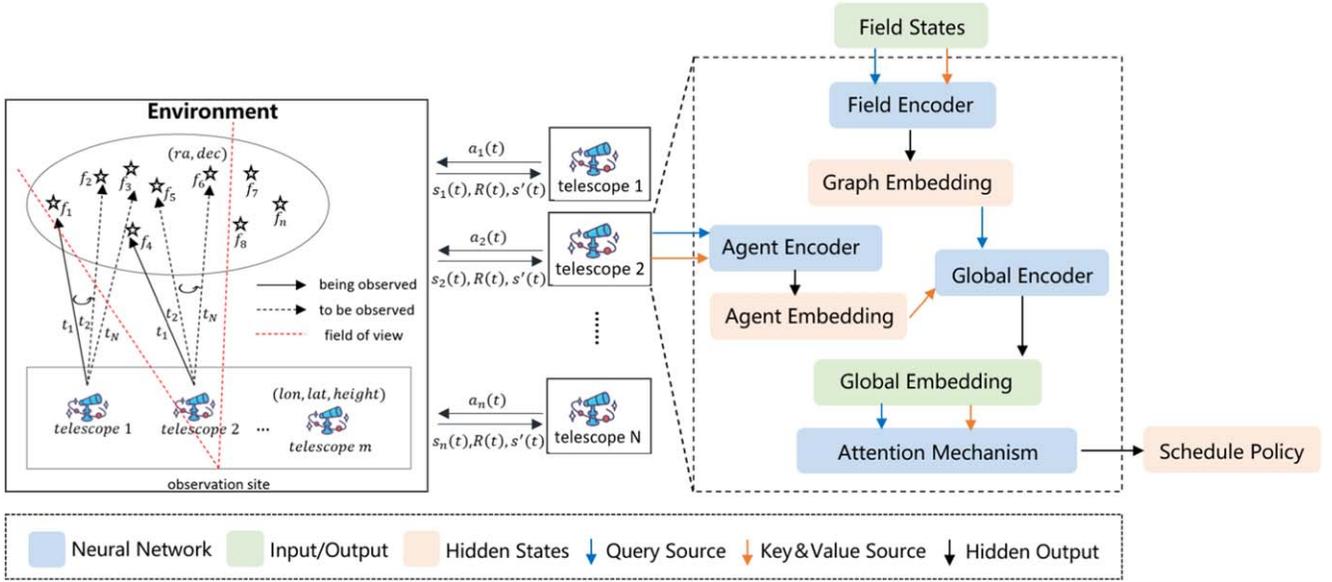

**Figure 4.** Architecture of the proposed intrasite observation scheduling scheme GRRIS for telescope array surveys. It mainly consists of a field encoder, an agent encoder, and a global decoder using an attention mechanism. The output is the final field selection policy, distributing observation tasks among different telescope agents.

$$q_i = \theta_{ig}^{d_k \times 2p} f_c^{2p}$$
$$k_{ij} = \theta_{ik}^{d_k \times p} f_j^p, \ j = \{1, 2, ..., n\}$$
$$v_{ij} = \theta_{iv}^{d_v \times p} f_j^p, \ j = \{1, 2, ..., n\}. \quad (5)$$

After obtaining the dimensions of the keys and values, $d_k$ and $d_v$ respectively, we calculate the compatibility of the query vector, which is linked to the agent, with all nodes,

$$u_{ij} = \frac{q_i^T k_{ij}}{\sqrt{d_k}}, \ j = \{1, 2, ..., n\}. \quad (6)$$

Afterward, the attention weights $w_{ij} \in [0, 1]$ can be obtained, which the agent assigns to node $j$ using a softmax function:

$$w_{ij} = \frac{e^{u_{ij}}}{\sum_{j'} e^{u_{ij'}}}, \ j = \{1, 2, ..., n\}, j' = \{1, 2, ..., n\}. \quad (7)$$

Utilizing the attention weights, the agent embedding can be generated as follows:

$$h_i = \sum_j w_{ij} v_{ij}. \quad (8)$$

### 3.4. Distributed Policy Network

The proposed model leverages probabilistic assignment and can be trained using a policy gradient. To compute the policy that assigns a specific node $j$ to an agent $i$, we first calculate the importance of each agent with respect to node $j$. For the agent $i$,

$$k'_{ij} = \theta_{ik'}^{d'_k \times p} f_j^p, \ j = \{1, 2, ..., n\}$$
$$q'_i = \theta_{iq'}^{d'_k \times d_v} h_i^{d_v}$$
$$u'_{ij} = \frac{q''^T_i k'_{ij}}{\sqrt{d'_k}}, \ j = \{1, 2, ..., n\}, \quad (9)$$

with $d'_k$ denotes the dimension of new keys, and the projection of embeddings back to $d'_k$ dimensions is achieved using parameters $\theta_{ik'}$ and $\theta_{iq'}$ in neural network architectures. For a new round of attention computation, after getting $u'_{ij}$, the results will be clipped into $[-C, C]$, with $C = 10$, followed by research (I. Bello et al. 2016), which can be adjusted according to the training data in different observation scenarios. The importance of node $j$ for agent $i$, denoted by $\text{imp}_{ij}$, can be obtained as

$$\text{imp}_{ij} = C \tanh(u'_{ij}), \ j = \{1, 2, ..., n\}. \quad (10)$$

Finally, we introduce softmax over all agents for every node to create a probabilistic assignment function that can then be optimized using the policy gradient

$$p_{ij} = \frac{e^{\text{jmp}_{ij}}}{\sum_a e^{\text{jmp}_{ij}}}, \ i \in \{1, 2, ..., m\}, j \in \{1, 2, ..., n\}. \quad (11)$$

The probability of agent $i$ selecting node $j$, $p_{ij}$ is calculated to estimate the probability of agents being granted the privilege. It should be noted that the parameters in the policy networks are dependent on the agents, not the nodes. This means that the method can be applied to a varying number of cities without needing to retrain the network. However, if the number of agents used changes, retraining would be necessary.

To demonstrate the advantage of RL, we try to reduce the reliance on domain-specific knowledge in our approach. In our design, the reward is simply the negative of the maximum observation cost among all agents, and all agents share it as a global reward. An important characteristic of this reward structure is its sparsity, as agents can only obtain the reward upon the completion of the entire astronomical observation task by all agents. As shown below, to estimate the parameters $\theta$ of our model, our aim is to maximize the cumulative reward obtained by the policy $\theta^* = \arg\max_\theta \mathcal{L}_R(\theta)$,

$$\mathcal{L}_R(\theta) = \mathbb{E}_{(G,m,n)\sim\mathcal{D}} \mathbb{E}_{\lambda \sim \pi(\theta)} R(\lambda)$$
$$= \mathbb{E}_{(G,m,n)\sim\mathcal{D}} \sum_\lambda \pi_\theta(\lambda|G, m, n) R(\lambda), \quad (12)$$

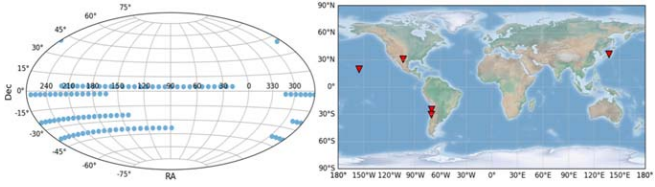

**Figure 5.** The distribution of observation sites and fields in data set generation. The blue dots are the sky regions where scheduled observations are needed in the sky survey mode of the telescope array. The R.A. range is [19°.26, 312°.50], while the range in decl. is [−89°.05, 76°.55]. We select five different observatories or telescopes from the Astropy package (A. M. Price-Whelan et al. 2018) to simulate the locations of the sites contained in a telescope array, which are marked by red triangles.

where $\mathcal{D}$ denotes the training set, and $\lambda$ denotes the assignment of fields to specific agents for astronomical observation. $R(\lambda)$ is the reward of assignment $\lambda$. $\pi_\theta(\lambda) = \prod_{j\in\{1,\ldots,n\}} p_{ij}$, which means that for an agent $i$, $\pi(\theta)$ is the distribution of all field assignments over parameter $\theta$. Agents share the cumulative reward to cooperatively minimize the maximum observation cost $R(\lambda)$. Therefore, after training the network parameters of GRRIS by the RL algorithm, it can find the optimal policy as well as the corresponding ultimate schedule effectively.

## 4. Experiments and Discussion

In this section, we conduct experiments on simulated data based on various real-world observation scenarios to investigate the effectiveness of GRRIS in terms of solution quality, computational speed, and scalability.

### 4.1. Experimental Setup

#### 4.1.1. Simulation Environment

We develop a Python simulator (Y. Zhang et al. 2023) to model telescope array observations, including the creation of scheduling blocks obtained from a long-term global scheduler, observation conditions, telescope equipment status, and observation fields to evaluate the effectiveness of our proposed scheduling algorithm under various settings. The simulator serves for both model training and evaluation. Note that the number of observation tasks is adjusted based on the telescope's exposure time and scheduling block duration from the global scheduler in each simulated scenario. All computations are conducted on a Ubuntu server with 4 Intel Xeon processors and a Tesla V100 GPU.

#### 4.1.2. Training Settings and Network Structure

We consider three real-world cases for model training with different numbers of telescopes (observation capacities) and tasks, namely, *Case 1* (three telescopes, 36 tasks), *Case 2* (five telescopes, 60 tasks), and *Case 3* (10 telescopes, 120 tasks). Note that in order to enrich the training samples and improve the generalization ability of the model, we randomly generate scheduling scenarios from a series of sky coordinates and observation site positions to simulate the site scheduler decision-making in the telescope array time-domain survey. To be specific, given the number of telescopes $m$ and observation fields $n$, positions of $n$ fields are randomly generated from 100 sky areas (shown in Figure 5), in R.A. and decl. range [19°.26, 312°.50] and [−89°.05, 76°.55], respectively. The positions of the site in the telescope array is randomly chosen from five real sites from the Astropy (A. M. Price-Whelan et al. 2018) library. Their details and geographical distribution are presented in Table 1 and Figure 5.

**Table 1**
Summary of Observation Site Information Used in the Experiments to Simulate the Composition of the Telescope Array

| Site | R.A. (deg) | Decl. (deg) | Altitude (m) |
|---|---|---|---|
| Subaru | −155.48 | 19.83 | 4139.00 |
| KAGRA | 137.31 | 36.41 | 414.18 |
| Cerro Tololo | −70.82 | −30.17 | 2215.00 |
| McDonald Observatory | −104.02 | 30.67 | 2075.00 |
| Paranal Observatory | −70.40 | −24.63 | 2669.00 |

**Table 2**
Parameter Settings in the Proposed Network Environment

| Parameter | Value |
|---|---|
| Length of a scheduling block | 30 minutes |
| Exposure time (exp) | 3 minutes |
| R.A. | [19°.26, 312°.50] |
| Decl. | [−38°.65, 40°.55] |
| Start time of the scheduling | 2015-06-17 4:00:00 |
| Learning rate | $1e-5$ |
| Batch size | 12 |
| Iteration | 3000 |
| Optimizer | Adam |

The same rules are used to generate the testing data set and training data set. During training, the data is randomly generated with each iteration. We generate 1000 instances for each problem scale of the testing data set. Table 2 presents the main simulation parameter and hyperparameter settings in the training. The scheduling block length and exposure time during training are 30 and 3 minutes, respectively. The beginning time of the scheduling blocks is 2015 June 17 at 4:00:00. The batch size is 12, and the learning rate is $1e-5$. Note that to respond promptly to special site observation challenges and enhance global survey efficiency, we include two redundant tasks in each block (thus totaling 36, 60, or 120 tasks). This allows for quick task substitution in case of unforeseen issues.

#### 4.1.3. Evaluation Metrics

Two metrics are used to evaluate the performance of algorithms.

1. *Observation cost:* As described in Section 2, it evaluates the observation quality of all scheduled tasks in the scheduling block. We evaluate the solution quality in terms of the maximum sum of observation cost (here calculated by airmass) among agents in the generated solution, and the standard deviation of the results of various test instances (*maxCost* and *std* for short, respectively).
2. *Computation time:* It means the average policy computation time of the scheduling algorithm (*time* for short).

#### 4.1.4. Baselines

Since there is no feasible method for intrasite observation scheduling in the telescope array survey environment, we try to use various linear programming, heuristic methods, and open-

Table 3
Performance Evaluation Based on Case 1, Case 2, and Case 3

| Method | Case 1 | | | Case 2 | | | Case 3 | | |
| --- | --- | --- | --- | --- | --- | --- | --- | --- | --- |
| | Max Cost | Std | Time (s) | Max Cost | Std | Time (s) | Max Cost | Std | Time (s) |
| Greedy | 18.32 | 0.72 | 4.03 | 19.33 | 0.59 | 11.96 | 19.56 | 0.66 | 50.19 |
| Gurobi | 22.33 | 6.41 | 0.86 | 38.43 | 14.15 | 58.67 | … | … | … |
| ORTools | 15.99 | 0.96 | 14.25 | 25.93 | 7.45 | 45.57 | 89.42 | 32.10 | 95.05 |
| GA | 32.72 | 11.20 | 828.81 | 44.45 | 15.17 | 1368.58 | 57.09 | 16.96 | 2492.36 |
| SA | 32.89 | 11.46 | 294.64 | 45.57 | 15.51 | 504.30 | 54.56 | 16.44 | 1043.71 |
| **GRRIS** | **14.22** | **1.86** | **0.25** | **18.63** | **3.36** | **0.50** | 39.91 | 7.59 | **0.57** |

**Notes.** Experimental results are evaluated in terms of the maximum observation cost, the standard deviation of observation cost between telescopes, and computation time. The bold parts highlight the advantages of the results of the GRRIS method we propose in this paper compared with other methods.

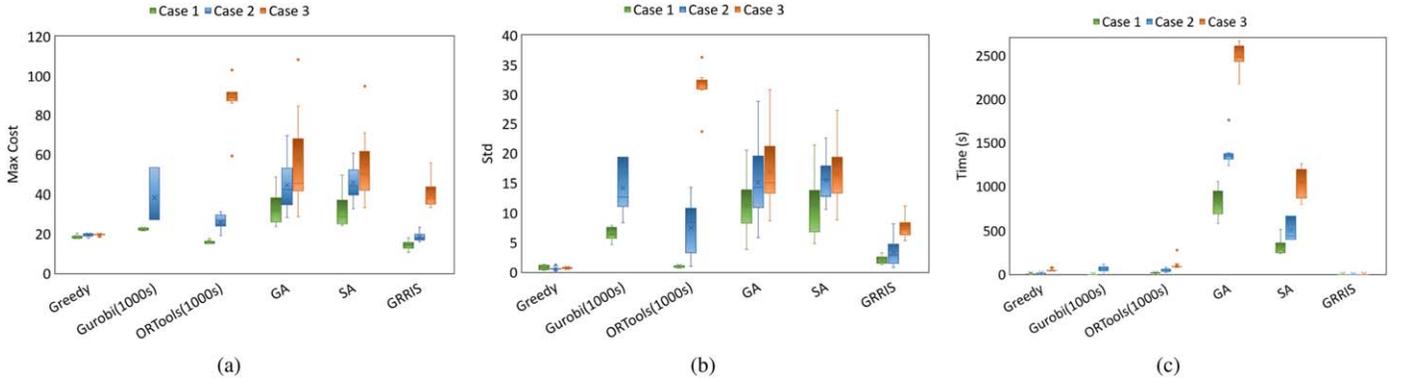

**Figure 6.** Comparisons of the observation cost, std, and the computation time for Case 1, Case 2, and Case 3 under different scheduling methods.

source optimization tool kits to apply to this problem. Five typical scheduling schemes are introduced for the fairness of performance comparisons.

*MiOCE.* The site scheduler follows a greedy policy, selecting observation tasks with the lowest immediate cost at each exposure time without considering global future consequences.

*Gurobi.* Gurobi (Gurobi Optimization 2014) has the potential to obtain an optimal solution for the observation scheduling problem, which can be formulated as an ILP problem. However, the observation scheduling problem has a vast search space that involves $mn^2$ binary variables under constraints, which demands a substantial amount of time to get the optimal solution. Here, we set a time limit of 1000 s to select the best solution among all feasible ones because excessively long computation times are impractical for real observation scheduling needs.

*ORTools.* ORTools[5] employs meta-heuristics, specifically local search, to enhance the initial solution iteratively until it reaches a local optimum. As the number of fields and telescopes increases, the efficacy of local search diminishes, and the computational cost increases substantially. So, we set ORTools to perform local search within a time limit of 1000 s.

*Genetic algorithm (GA).* We implement a scheduling method based on a GA. This population-based meta-heuristic algorithm employs natural selection to iteratively merge the fittest individuals from an initial population, generating new solution generations. Utilizing a fitness function (similar to our proposed observation cost) to assess solutions and incorporate crossover and mutation, it converges toward an optimal solution.

*Simulated annealing (SA).* We also implement a benchmark using SA. To avoid local optima, a probabilistic acceptance criterion is utilized to explore the search space and gradually converge toward the global optimum solution. New solutions are generated by perturbing the current solution, and their acceptance is determined by a gradual decrease in temperature.

### 4.2. Comparisons of Solution Quality

As shown in Table 3, we train three models based on three cases of different scales, and conduct testing on the same-scale simulation cases. In order to ensure a fair comparison, we use one thread on the CPU to execute GRRIS, similar to how ORTools searches for the solution. However, GRRIS has the advantage of being able to leverage tensor operations, allowing the assignment part to be executed in parallel, which proves to be more efficient when run on the GPU.

The solution quality of the scheduling is compared with solutions from other solving methods in terms of maximum observation cost and standard deviation of observation quality between telescopes, as illustrated in Figure 6 and Table 3. Values are obtained from the average of 100 test instances. The results verify that GRRIS performs better than all other baselines in Case 1 and Case 2, and is second only to the greedy method in Case 3. In larger instances, ORTools perform poorer than other baselines and has stronger uncertainty. Gurobi is unable to generate a solution in Case 3 within the time limit of 1000 s, which also verifies the complex high-dimensional state space in the telescope observation scheduling problem. Meanwhile, the increase in *std* is relatively stable as the size of the problem scale increases, indicating that the scheduling solution generated by GRRIS results in a more balanced workload and observation effect between telescopes. The performance of Gurobi and GA also degrades faster with respect to the problem scale increases. This is because GRRIS enhances rewards through collaborative learning from interactions with the environment, and exhibits greater flexibility in addressing high-dimensional state spaces and dynamic conditions. Conversely, traditional tools typically demand predefined constraints and objective functions, making them less suitable for intricate and dynamically scheduling challenges. By testing different instances, it can be seen in Figure 6 that GRRIS shows more robustness, while ORTools and GA are more likely to produce abnormal scheduling results with large differences from the mean.

---

[5] https://developers.google.com/optimization/

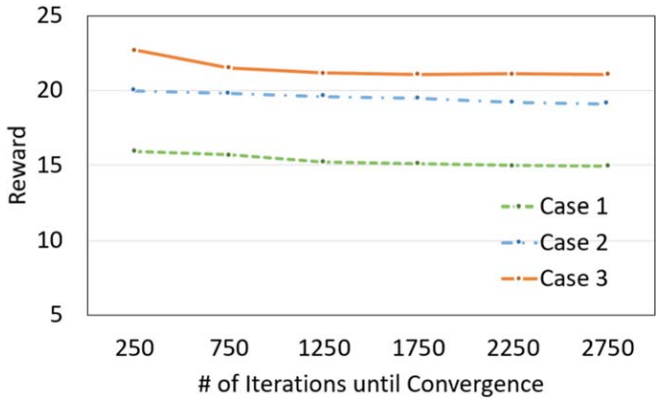

**Figure 7.** Convergences of the rewards (maximum astronomical observation cost among all telescope agents) of different test scenarios.

### 4.3. Comparison of Performance

We can observe from Table 3 that compared with the baseline methods, GRRIS consistently achieves the shortest execution duration to obtain a solution. Moreover, it provides efficient solutions in under a second for various problem sizes, with the potential for further improvement using GPUs. In comparison, methods based on linear programming and heuristics require minutes of solving time (while GA can take up to hours). Meta-heuristic algorithms, for instance, iteratively enhance solutions until they converge to local optima. As the search space expands, their performance progressively diminishes, coupled with a notable rise in execution time.

Experimental results show that GRRIS's neural network structures exhibit high computational efficiency during the inference process and can acquire an effective policy with reduced computational cost. In real surveys, where each exposure is only a few minutes, lengthy task allocation times can significantly waste observation resources. For example, in Case 3, the quality of the solution of GRRIS is reduced compared with that of the greedy algorithm, but the reasoning speed is 1% of that of the greedy algorithm, so it is more desirable in the actual observation project. Moreover, a real-time task allocation strategy enables the telescope array to operate with greater flexibility, especially in response to emergency situations.

### 4.4. Convergence Analysis

Figure 7 demonstrates the rewards obtained by different telescope agents at various training iterations. The rewards of the agents all converge after around 2500 iterations. The successful convergences of the agents demonstrate their ability and effectiveness to achieve equilibrium in the mixed cooperative-competitive environment in various scales of site scheduling cases. The converged rewards (the maximum observation cost of agents) increase with the scale. This is partly because a larger telescope array involves scheduling a greater number of telescopes, leading to an expansion of the task state space and an increase in the complexity of scheduling. Simultaneously, as the number of telescopes increases, the collaboration and implicit competition between agents become more complex, and a more fine-grained scheduling strategy is needed to balance resource utilization and task allocation among agents. Therefore, it incurs more difficulty for the scheduling model training accordingly. Moreover, Case 3 demonstrates lower reward fluctuations during training, likely attributed to the abundance of training samples in larger-scale scheduling problems. This facilitates a comprehensive understanding of task features and strategies, decreasing training randomness and volatility. Because achieving satisfactory rewards in the initial stages of DRL is often challenging due to the agent's need to explore the environment, the complexity of learning from sparse or delayed rewards, and the time required to optimize learning parameters and policies. However, in larger-scale scheduling problems, the abundance of training samples can lead to better rewards by providing more opportunities for the agent to learn effective strategies and refine its decision-making process.

### 4.5. Generalization on Larger-scale Problems

GRRIS enables training and testing on graphs of different sizes using the same model parameters. To evaluate the generalization, we apply the trained model to larger-scale, unseen problems in each of the three designed cases. In most time-domain surveys (e.g., X. Yuan et al. 2014, 2020; E. C. Bellm et al. 2018; Ž. Ivezic et al. 2019; Y.-P. Chen et al. 2023), the exposure time is less than 60 s. As shown in Tables 4, 5, and 6, to explore telescope array time-domain surveys with longer single exposures, here we set the exposure times to 1, 2, and 3 minutes, respectively. And we design test instances of observation tasks at various scales based on exposure times in different cases. As the number of telescopes and tasks increases, Gurobi fails to produce a feasible solution within a reasonable time, so it is excluded from the table. As the size of the scheduling problem grows, while GRRIS may rank second to the greedy algorithm in specific instances, the increase in observation cost remains consistently stable. GA, SA, and ORTools, however, show more volatile results. ORTools exhibits its superior solution quality with just three telescopes and 45 tasks. In addition, the computation time grows dramatically for GA and SA, while GRRIS can obtain results in an average time of 0.58 s with little difference under a variety of observation configurations, indicating its ability for quasi-real-time scheduling. The greedy algorithm achieves optimal solutions with a computational complexity of $O(Nn\lg n)$, incurring significant computational costs due to the extensive search space. In contrast, GRRIS effectively learns a general policy representation. While this representation may not always yield optimal solutions and can, in some cases, result in poorer solutions compared to other algorithms, it generally excels in generating higher-quality solutions in subsecond.

**Table 4**
Generalization Results on Field Numbers {45, 90, 135, 180} with exp = 1 by the Model Trained on Instances with Three Agents and 36 Fields

| Method | fnum = 45 | | | fnum = 90 | | | fnum = 135 | | | fnum = 180 | | |
|---|---|---|---|---|---|---|---|---|---|---|---|---|
| | maxCost | Std | T (s) | maxCost | Std | T (s) | maxCost | Std | T (s) | maxCost | Std | T (s) |
| Greedy | 23.61 | 0.18 | 8.77 | 46.82 | 0.14 | 34.83 | 67.06 | 0.21 | 79.75 | 87.34 | 0.31 | 218.73 |
| GA | 51.32 | 20.58 | 1109.37 | 81.03 | 29.11 | 3025.33 | 107.37 | 36.97 | 4637.87 | 157.23 | 55.21 | 6389.38 |
| SA | 38.97 | 12.25 | 525.49 | 95.34 | 38.48 | 1509.46 | 172.26 | 73.31 | 2014.77 | 193.53 | 75.58 | 2125.98 |
| ORTools | 22.17 | 0.83 | 20.31 | 46.02 | 6.12 | 93.38 | 92.94 | 22.45 | 170.85 | 132.63 | 64.63 | 298.67 |
| **GRRIS (3, 36)** | **25.72** | **3.17** | **0.35** | **51.08** | **6.84** | **0.58** | **75.67** | **8.45** | **0.61** | **101.55** | **12.01** | **0.52** |

Table 5
Generalization Results on Field Numbers {75, 100, 125, 150} with exp = 2 by the Model Trained on Instances with Five Agents and 60 fields

| Method | fnum = 75 | | | fnum = 100 | | | fnum = 125 | | | fnum = 150 | | |
|---|---|---|---|---|---|---|---|---|---|---|---|---|
| | maxCost | Std | $T$ (s) | maxCost | Std | $T$ (s) | maxCost | Std | $T$ (s) | maxCost | Std | $T$ (s) |
| Greedy | 23.04 | 0.51 | 20.54 | 38.91 | 0.55 | 34.45 | 39.52 | 0.83 | 103.65 | 45.29 | 0.77 | 117.62 |
| GA | 60.63 | 22.29 | 2484.39 | 39.26 | 6.98 | 1889.06 | 75.99 | 25.59 | 3826.49 | 178.52 | 65.07 | 4481.54 |
| SA | 51.23 | 16.86 | 947.67 | 45.77 | 3.72 | 782.88 | 100.48 | 33.91 | 1547.74 | 151.61 | 54.72 | 1942.96 |
| ORTools | 30.96 | 9.69 | 41.83 | 52.95 | 23.28 | 64.67 | 89.71 | 40.48 | 151.91 | 119.76 | 53.12 | 233.26 |
| **GRRIS (5, 60)** | **26.51** | **4.74** | **0.57** | **38.42** | **6.56** | **0.63** | **49.77** | **6.19** | **0.58** | **71.48** | **11.72** | **0.72** |

Table 6
Generalization Results on Field Numbers {100, 140, 170, 200} with exp = 3 by the Model Trained on Instances with 10 Agents and 120 Fields

| Method | fnum = 100 | | | fnum = 140 | | | fnum = 170 | | | fnum = 200 | | |
|---|---|---|---|---|---|---|---|---|---|---|---|---|
| | maxCost | Std | $T$ (s) | maxCost | Std | $T$ (s) | maxCost | Std | $T$ (s) | maxCost | Std | $T$ (s) |
| Greedy | 15.63 | 0.89 | 71.58 | 21.29 | 0.94 | 89.36 | 25.95 | 1.23 | 133.94 | 31.61 | 1.31 | 201.46 |
| GA | 46.36 | 14.38 | 2097.48 | 56.20 | 16.05 | 2967.60 | 79.60 | 22.31 | 3814.51 | 81.72 | 22.79 | 4522.44 |
| SA | 50.09 | 13.54 | 860.76 | 71.92 | 21.40 | 1008.63 | 98.90 | 27.85 | 1191.16 | 89.42 | 25.27 | 1425.68 |
| ORTools | 55.88 | 24.37 | 102.18 | 115.93 | 52.14 | 180.59 | 154.05 | 69.43 | 242.79 | 184.23 | 88.43 | 325.73 |
| **GRRIS (10, 120)** | **39.33** | **5.42** | **0.48** | **48.68** | **15.36** | **0.59** | **56.60** | **19.49** | **0.60** | **78.13** | **20.23** | **0.75** |

## 5. Conclusion and Future Work

The future will see new imaging and spectroscopy surveys on an unprecedented scale. Collaborative observations using telescope arrays have led to new patterns of observation, facilitating more frequent and extensive observations. The unique challenges of the astronomical survey observation, such as vast search spaces, limited labeled data, and computational costs, motivate the proposal of GRRIS, a real-time observation scheduling scheme for practical site schedulers in a distributed telescope array. GRRIS uses the GNN to extract spatially informed features and optimize agent-node allocation policies via MARL, leading to efficient scheduling.

Numerical simulations using real-world scenarios demonstrate GRRIS's superior performance in solution quality and computational time compared to several competitive scheduling schemes. Compared with the greedy algorithm, GRRIS improves solutions by 22% and 4%, respectively, in Case 1 and Case 2, while offering a 1% solution time in Case 3, favoring quasi-real-time astronomical scheduling. GRRIS also displays robustness and scalability, capable of handling larger-scale instances and various input configurations in an average of 0.58 s, without additional training or tuning.

Our proposed GRRIS is a novel scheme to optimize the observation strategy for telescope arrays. In the future, we will explore dynamic observation scheduling in more complex scenarios, such as the emergence of transient source tracking targets, heterogeneous telescopes within the observation site (oriented to different bands or targets), field prioritization based on science goals, etc. A more refined scheduling model and the balance of multiobjective optimization need to be explored. And we would further use it to control practical telescope arrays, such as the Sitian project (J. Liu et al. 2021).

### Acknowledgments

This work is financially supported by the National Key Research and Development Program of China No. 2023YFA1608301, and the National Natural Science Foundation of China (NSFC) No. 12133010 and No. 12273025. Data resources are supported by the China National Astronomical Data Center (NADC), the Chinese Academy of Sciences (CAS) Astronomical Data Center, and the Chinese Virtual Observatory (China-VO).

### Data Availability

The data behind the figures and the data samples used in this study are available via doi:10.12149/101472.

### Conflicts of Interest

The authors declare no conflicts of interest.

## ORCID iDs

Yajie Zhang  https://orcid.org/0000-0003-2976-8198
Yi Hu  https://orcid.org/0000-0003-3317-4771